\documentclass[%
aip,
rsi,%
amsmath,amssymb,
reprint,%
]{revtex4-1}

\usepackage{graphicx}
\usepackage{dcolumn}
\usepackage{bm}

\begin{document}
\preprint{AIP/123-QED}	

	\title[Advancing Fourier: space-time concepts in ultrafast optics, imaging and photonic neural networks]{Advancing Fourier: space-time concepts in ultrafast optics, imaging and photonic neural networks}
	
	\author{Luc Froehly, Fran\c{c}ois Courvoisier, Daniel Brunner, Laurent Larger, Fabrice Devaux, Eric Lantz, John M. Dudley, Maxime Jacquot$^{*}$}
	\affiliation{D\'{e}partement d'Optique P. M. Duffieux, Institut FEMTO-ST,  Universit\'e Bourgogne Franche-Comt\'e CNRS UMR 6174, Besan\c{c}on, France.
	       \\* Corresponding author: maxime.jacquot@univ-fcomte.fr}
	
	

%
%
%
%
%
%
	
	\date{\today}

%
%
%
%






\begin{abstract}
The concepts of Fourier optics were established in France in the 1940s by Pierre-Michel Duffieux, and laid the foundations of an extensive series of activities in the French research community that have touched on nearly every aspect of contemporary optics and photonics.  In this paper, we review a selection of results where applications of the Fourier transform and transfer functions in optics have been applied to yield significant advances in unexpected areas of optics, including the spatial shaping of complex laser beams in amplitude and in phase, real-time ultrafast measurements, novel ghost imaging techniques, and the development of parallel processing methodologies for photonic artificial intelligence. 
\end{abstract}


\maketitle

\section{Introduction}

The description of light as an electromagnetic wave was developed by Maxwell in the 1860s, and led to a fundamental change in the way in which optical phenomena were studied and understood.  With the oscillatory nature of light established, it became possible to apply new mathematical tools to quantify experiments, and it was not long before the concepts of wave decomposition were applied by Rayleigh and Abbe to the study of resolving power \cite{Hawkes-1997}.  More rigorous and systematic approaches to applying the tools of Fourier analysis in optics were developed in the 1930s, and in 1946, the first complete textbook on what we now describe as Fourier optics was published in French by Pierre-Michel Duffieux \cite{Duffieux-1946}. 

Although this work was very influential in the French community, Duffieux’s contributions were largely unappreciated on the international stage until 1959 when Born and Wolf drew attention to them in their seminal work { \em Principles of Optics} \cite{Born-1959}. Even so, Duffieux's role in developing these ideas is often forgotten because  Duffieux’s book was only translated in 1983 \cite{Duffieux-1983}, long after many excellent and independent treatments of Fourier optics had appeared in English \cite{Goodman-1968, Papoulis-1968,Gaskill-1978}. Yet Duffieux's neglect is undeserved and unfortunate because, as Emil Wolf himself has stated, ``Pierre-Michel Duffieux deserves to be remembered as the true originator of the field of Fourier Optics.’’ \cite{Hawkes-1997}

Moreover, within the optics laboratory that he founded in Besan\c{c}on in France, Duffieux had tremendous influence on generations of researchers and students.  Indeed, work in Besan\c{c}on in the 1960s led to some of the first studies of holography and spectral interferometry \cite{Froehly-1967,Froehly-1973}, with the results in Ref. \cite{Froehly-1973} being especially important in transferring impulse response and transfer function concepts from the spatial to the temporal domains. As we shall see below these results have seen important application in areas such as ultrafast pulse measurements, and the concept of spectral interferometry itself is now ubiquitous in many areas of optics, notably in the field of frequency-domain optical coherence tomography \cite{Leitgeb-2003}. Other work in the 1970s and 1980s saw the development of novel proposals to apply Fourier techniques to the shaping of ultrashort laser pulses \cite{Colombeau-1976,Froehly-1981}, as well to optical information processing and handwriting recognition \cite{Vienot-1976}.

Many of these early ideas, however, have seen widespread application only recently with technological developments in sources and instrumentation, but Duffieux’s legacy continues  strongly in Besan\c{c}on with active research programmes applying Fourier concepts to many state-of-the-art applications. In this paper, we present an overview of our recent work in several different areas, including the spatial shaping of complex laser beams both in amplitude and in phase, measurement of the temporal properties of ultrafast light fields, novel ghost imaging techniques,  and the development of parallel processing methodologies for photonic artificial intelligence. 

The intention of this paper is to showcase a selection of this work, providing a contemporary overview of research from the laboratory that Duffieux himself established in order to develop experimentally the ideas that he had earlier described mathematically.  We anticipate that the main impact of this paper will be to show how basic and general concepts in optical physics can find wide and often unexpected applications, and in fields unanticipated by those who initially developed them. In what follows, the treatment of these different topics is largely self-contained, but it is our hope that they provide a clear demonstration of how Duffieux's pioneering ideas have now become universal in many central areas of today's optics and photonics.

\section{Spatial shaping of broadband beams}
Spatial beam shaping has now become an indispensable tool in many areas of research in optics, with much of its recent success being based on the development of convenient techniques for the design and tailoring of specific spatial beam properties by manipulating the amplitude and phase of spatial frequency components. In this section, as a particular example, we focus on spatially-tailored Bessel beams which have found wide use in a number of applications, including microscopy \cite{Fahrbach-2010}, optical trapping \cite{Garces-Chavez-2002}, and laser material processing \cite{Stoian-2018}. A particularly important application field is ultrashort pulse laser ``stealth dicing'' of transparent materials such as glass which enables for instance cutting coverglass for consumer electronics at very high speed (up to meters per second) \cite{Meyer_2019,Courvoisier_2016}. The Bessel beam transverse profile consists of a strongly-localized central ``hot spot'' surrounded by a large number of concentric circular rings. The central lobe possesses an extremely high aspect ratio maintained over propagation distances exceeding the Rayleigh range (of a conventional Gaussian beam with the same transverse extent) by orders of magnitude  \cite{Durnin-1987,McGloin-2005,Bock-2012}. 

For many applications, it is desirable to generate Bessel beams with a broadband spectrum while preserving the same lobe size (and the position of the intensity zeros in the profile) – over the full spectral extent.  These applications include those such as material processing using ultrashort pulses, Optical Coherence Tomography (where the highest imaging resolution is obtained for the broadest spectrum) and the generation of ``X-wave'' structures \cite{Saari-1997, Zamboni-Rached-2004}.  Several groups have previously investigated broadband, or ``white'' Bessel beams, but the full superposition of the different frequency components was effective only on the optical axis. Off-axis, the fringes were polychromatic \cite{Fischer-2005}.

In this section, we describe how the Fourier formalism allows us to shape Bessel beams out of broadband laser sources. 
Particularly useful is the property of convex lenses to be perfect ``Fourier-Transformers'' in the paraxial regime, as was pointed out by Duffieux. Indeed, if the scalar field amplitude in the object focal plane of a lens of focal length $f$ is $A(X,Y)$, then the amplitude in the image focal plane is:
\begin{equation}
 B(x,y) =  \iint{A(X,Y) e^{-\imath 2\pi \frac{(Xx+Yy)}{\lambda f}}dXdY}
 \label{fourierplane}
\end{equation}

\noindent In other words, the latter field $B$ is the perfect Fourier-transform of $A$ with $B(x,y)=\tilde{A}(\nu_x,\nu_y)$, where the spatial frequencies are defined as $\nu_x =x / (\lambda f)$,$\nu_y = y/(\lambda f)$   We note that placing a second lens in confocal configuration (i.e. we have a $4f$-system) now implements a second Fourier transformation, which gives in the image focal plane of the second lens the initial amplitude $A(X,Y)$. In the common focal plane,  ``Fourier filtering'' can easily be implemented , as we will use later here.

An ideal Bessel beam has a cylindrically-symmetric  transverse intensity profile of the form $|J_0(2 \pi \nu_0 r)|^2$, with $r$ the radius  in cylindrical coordinates, and where $\nu_0 = \frac{\sin \theta}{\lambda}$ is the spatial frequency with $\theta$ the {\it cone} angle of the Bessel beam at the wavelength $\lambda$. The Fourier transform of a Bessel beam is a ring of radius $\nu_0$, experimentally observed in the Fourier plane as a ring of radius $r_0=\lambda f \nu_0 = f \sin\theta$. Creating a broadband Bessel beam requires that this intensity distribution is invariant with wavelength, which is equivalent to requiring that the radial spatial frequency $\nu_0$ must also be invariant with wavelength. Hence, $\sin \theta(\lambda)$ should be proportional to $\lambda$.

In practice, reflective or refractive  axicons commonly used to generate Bessel beams cannot produce broadband (white) Bessel beams as defined above \cite{Boucher-2018, McLeod-1954}. The reason is that the phase imposed by such optical elements generates, at zeroth order in dispersion, a fixed cone angle $\theta$ (variation of spatial frequency inversely with wavelength.), such that the surrounding lobes are colored \cite{Fischer-2005}. 
Methods based on spatial light modulator (SLM) involving a reference wave are also limited because of the necessary compensation of the tilt with a prism, where the actual refractive index largely varies over the visible spectrum \cite{Leach-2006}. In contrast, we seek a method where the phase pattern is spatially independent of wavelength.

We use an approach derived from Lord Rayleigh's pioneering work \cite{Rayleigh-1889} where he was able to produce achromatic interference fringes using a Lloyd's mirror configuration combined with a diffraction grating together with a broadband light source. Specifically, we apply to the incoming broadband beam a phase mask in the form $\phi(r) = \frac{2\pi}{\lambda}T\delta n$ where $\delta n \sim J_0(2\pi \nu_0 r)$ is the local index variation, $T$ is the optical path length over the liquid crystal film thickness. For sufficiently small phase shifts, the amplitude at the output of the phase mask is on the form: $A(r) = A_0 (1+ J_0(2\pi \nu_0 r))$. Fourier filtering allows for selecting the second term. Indeed, a first convex lens realizes the Fourier transform of the beam after its reflection on the SLM. The intensity pattern in the Fourier space consists of a point (first term) at the center of an annulus (second term). A beam block at the center and an iris enable selecting only the annulus. A second Fourier transform is made using another lens to obtain a Bessel beam with fixed spatial frequency $\nu_0$ as desired. The diffraction grating-like structure produced here does introduce the necessary linear dependence of the $\sin \theta$ with the wavelength.  We note that in this approach, the impact of wavelength is limited only to the diffraction efficiency, which varies smoothly over the spectrum.
\begin{figure}[t]
	\includegraphics[width=0.48\textwidth]{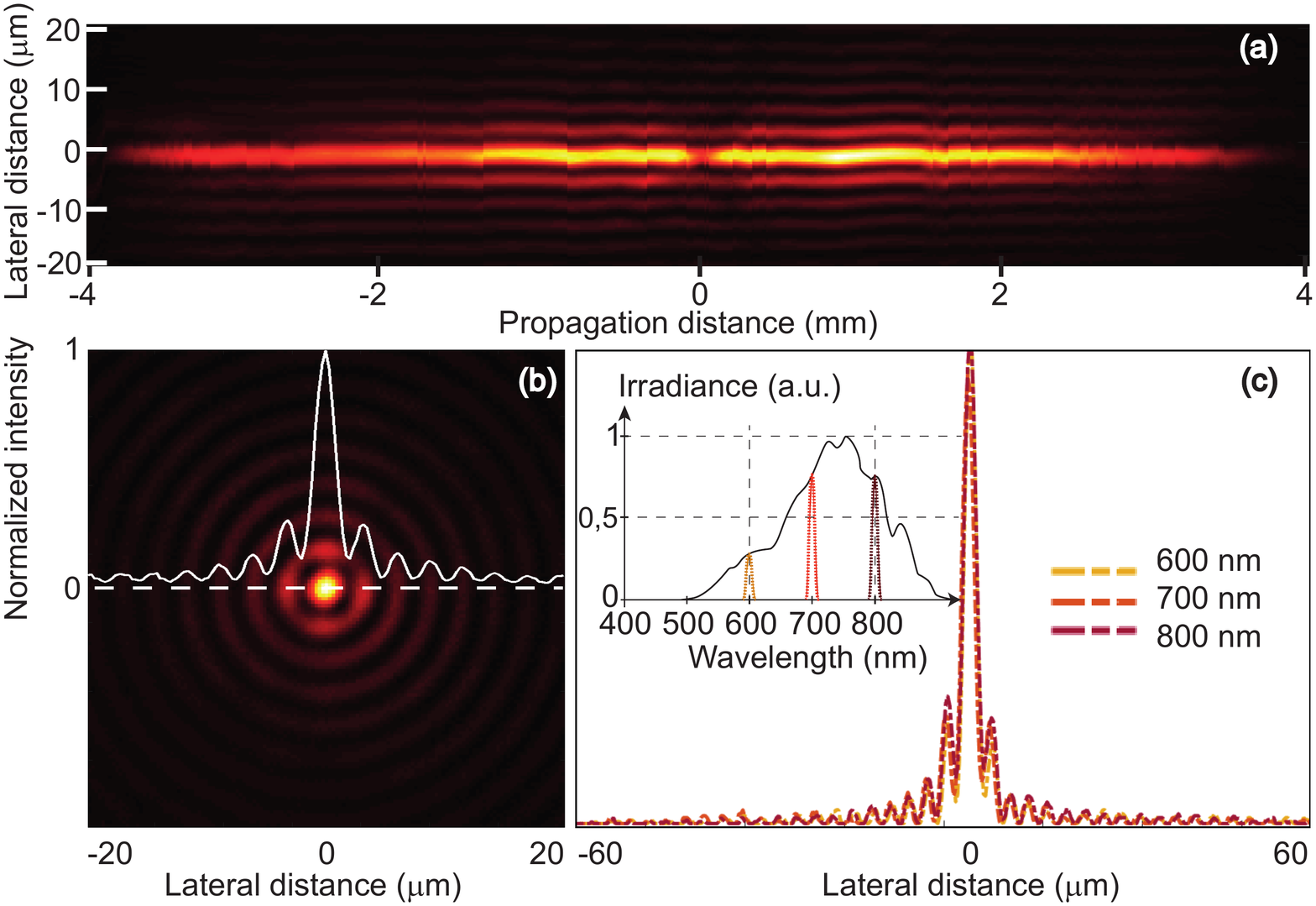}
	\caption{\label{fig:Bessel} (a) Intensity distribution of the Bessel beam with the full supercontinuum spectrum. The longitudinal propagation distance $z$ is measured from the image of the SLM mask in $z=0$. 
		(b) Transverse $(x,y)$ cross-section of the Bessel beam at position $z=-0.66~ mm$. The gray solid line shows a horizontal line-cut through the intensity transverse profile.
		(c) Normalized intensity profiles obtained for three narrow-band spectra centered respectively at 600, 700 and 800~nm wavelength.(inset) Solid black line shows the full supercontinuum spectrum. The three narrowband spectra used in (c) are also plotted.
}
	\end{figure}

Experimentally, we generate broadband Bessel beams using a supercontinuum source laser (LEUKOS SMHP-40-2-B) and a Spatial Light Modulator (SLM Hamamatsu X13138, with 1280~$\times$~1024 pixels). The source spectrum is freely adjustable in terms of central wavelength and bandwidth via translating filters (LEUKOS BEBOP). The phase mask is illuminated under normal incidence using a cube to separate input and output optical paths. A 2f-2f optical arrangement ($f_1$=200~mm and $f_2$ = 9.5~mm) is used to de-magnify the Bessel beam. In the Fourier plane of the first lens, we select the Bessel beam Fourier transform, {\it i.e.} a ring, from spurious orders of diffraction. We image the Bessel beam produced using a microscope objective, a relay lens with overall magnification factor of 10 and a CMOS camera mounted on a motorized translation stage as in reference \cite{Froehly2014}.

We remark that in the focal plane of a lens that materializes the Fourier plane, the radius of this ring is expressed as
$R={f~\nu_0\lambda}/{\sqrt{1-(\lambda\nu_0)^{2}}}$ where $f$ is the focal length of the lens that produces the Fourier transform.
Hence, each individual wavelength creates a Bessel ring with a different radius $R$, even if all have the same spatial frequency $\nu_0$.

We generate a Bessel beam with more than 200~nm bandwidth, measured at Full Width at Half-Maximum (FWHM), centered at 750~nm. The supercontinuum spectrum is shown in the inset of Fig.\ref{fig:Bessel}(c) (solid black line). The effective spatial frequency of the Bessel beam is $\nu_0=100$~mm$^{-1}$ (using the de-magnification factor of $\frac{1}{21}$ between the SLM plane and its image at the focal plane of the second lens). In Fig.\ref{fig:Bessel}, we plot the evolution of the measured intensity distribution with propagation distance. In Fig.\ref{fig:Bessel}(a), we show the experimental $(x,z)$ cross-section of the beam intensity recorded using the full input spectrum.  A  7~mm long Bessel beam is generated with a central lobe size of 10~$\mu$m. The contrast of the lobes is high ($\sim$~0.8 to 0.9), even if the input spectrum spans over more than 200~nm. This is even more apparent from the transverse cross sections shown in  Fig.\ref{fig:Bessel}(b).

To demonstrate the invariance of the position and contrast of the Bessel beam lobes with wavelength, we have independently recorded, using the same phase mask, the cross section corresponding to three narrowband input spectra, centered at 600, 700 and 800~nm with $\sim$~5~nm bandwidth FWHM. We plot in Fig. \ref{fig:Bessel}(c) the normalized intensity cross sections. It is apparent that the lobes are quasi identical, even up to $\sim$~60~$\mu$m from the optical axis. Small chromatic effects, in the noise limit, remain in our experiment. However, we stress that this is due to the imaging lenses which are not achromatic and not to the generation of the Bessel beam itself. The chromatism induces a variation in the magnification factor of about 1\% over the 200~nm bandwidth (experimentally characterized). This could be avoided using an imaging system fully based on reflective optics.

In conclusion, the application of the classical tools of Fourier optics has allowed us to generate a Bessel beam with more than 200~nm spectral bandwidth, spatially extending over more than 7~mm with a central lobe diameter of 10~$\mu$m FWHM. This corresponds to an aspect ratio of 1:700. We believe this will open new perspectives for applications of Bessel beams to microscopy, optical coherence tomography and ultrafast physics \cite{Blatter-2011,Yu-2015}.

\section{Real-time measurements of non-repetitive signals}

The concepts of Fourier optics are widely applied to manipulate and measure the temporal properties of ultrafast light fields by exploiting the analogy between diffraction in space and dispersion in time. The power of this ``space-time analogy'' was appreciated in numerous early works \cite{Froehly-1981}, and practical techniques for Fourier-transform pulse shaping are now widespread \cite{Weiner-2009}.  Recent years have seen particular progress applying space-time concepts to develop experimental techniques to measure ultrafast spectral and temporal instabilities in real-time, and such methods are rapidly becoming standard in the characterization of a wide range of nonlinear propagation phenomena \cite{Solli-2007,Solli-2012,Wetzel-2012,Suret-2016,Narhi-2016,Tikan-2018,Ryczkowski-2018}.      

The first technique of this sort that saw widespread use was the Dispersive Fourier Transform (DFT) which enables the convenient measurement of real-time spectra at MHz repetition rates \cite{Goda-2013}. This principle of the DFT uses the fact that a pulse propagating in a linear dispersive medium evolves towards its Fourier transform with sufficiently large value of quadratic spectral phase (from group velocity dispersion).  This is exactly equivalent to the spatial phenomenon where far-field diffraction from an arbitrary spatial mask yields the corresponding spatial Fourier transform.  

In particular, consider a pulse propagating in an optical fiber with temporal envelope $A(z,t)$, where $z$ is distance and $t$ is co-moving time (i.e. in a frame of reference moving at the pulse group velocity).  If  $A(z,t) \leftrightarrow \tilde{A}(z,\omega)$ defines the corresponding spectral amplitude through a Fourier transform, then the effect of purely dispersive propagation on an incident field [with spectrum $\tilde{A}_0(\omega) =\tilde{A}(z=0,\omega)$] is to introduce a quadratic spectral phase:
\begin{equation}
\tilde{A}(z,\omega) =  \tilde{A}_0(\omega)\, e^{\frac{i}{2}\beta_2\,z\,\omega^2}.   
\end{equation}
\noindent If we now apply the inverse Fourier transform to determine the corresponding time-domain field and assume a large value of dispersion $\beta_2\,z$, then we can use the stationary phase approximation or steepest descent method to show the mathematical equivalence $|A(z,t)|^2 \propto |\tilde{A}_0(t/\beta_2\,z)|^2$ which also provides the frequency to time scaling $\omega = t/(\beta_2\,z)$ \cite{Grigoryan-2000}. This relation  describes how the timebase of the temporally dispersed (stretched) pulse can be related to the optical frequency.

A typical DFT experimental setup considers a fluctuating field typically of 10's of picosecond duration that is injected at low power into a length of optical fibre such that at the fibre output it has stretched to a duration of 5-10~ns. It is this stretched field that has the functional form of the corresponding power spectrum, and can be conveniently measured using a suitable high speed detector and oscilloscope system (with typically 10s of GHz bandwidth).  The DFT technique attracted particular attention in 2007 when it was used to reveal the presence of extreme value ``rogue wave'' fluctuations in the spectra of broadband supercontinuum light generated in photonic crystal fiber \cite{Solli-2007}.   

\begin{figure}[ht]
	\includegraphics[width=0.5\textwidth]{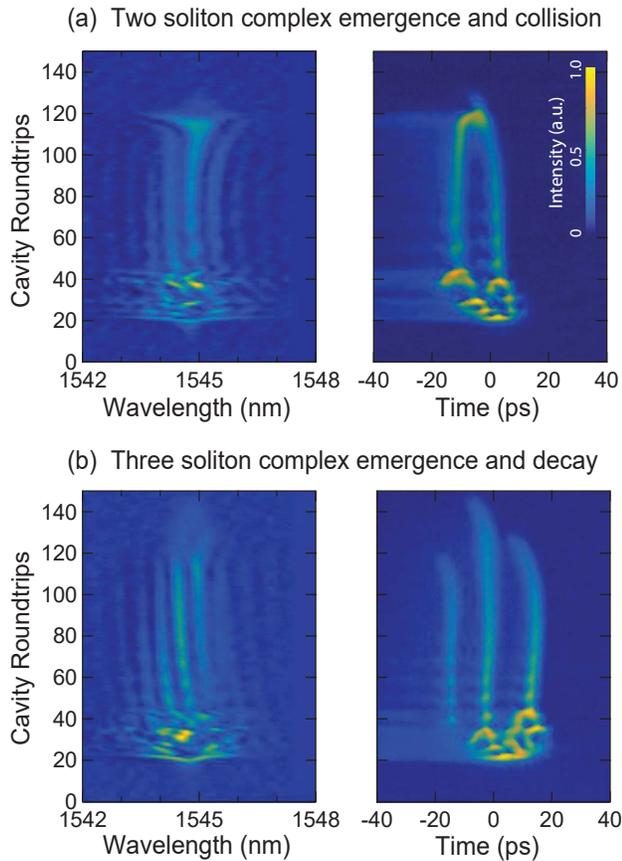}
	\caption{Typical results from simultaneous DFT and time-lens characterization of fibre laser start-up dynamics (see also Ref.~\cite{Ryczkowski-2018}). (a) shows emergence of a 2-soliton state where the two components separate then merge before sudden decay. (b) shows a 3-soliton state where the components evolve independently before decay. Left and right columns show  the measured spectra and temporal intensity profiles respectively.}
	\vspace{-0.5cm}
\end{figure} 

Complementing the application of the space-time analogy to the development of DFT to measure real-time spectra, parallel work has developed time-domain systems equivalent to the thin lens to allow direct detection of transient ultrafast temporal processes \cite{Froehly-1981,Kolner-1989}. The concept of the ``time lens'' takes an incident picosecond field, stretches it in a length of dispersive fiber, applies a quadratic temporal phase, and then stretches it again in a second length of dispersive fiber.  By matching the dispersive fiber lengths to the magnitude of the applied phase \cite{Salem-2013}, it is possible to create an equivalent temporal imaging system that  magnifies picosecond pulses to the nanosecond domain whilst maintaining their intensity profile.  Using a suitable high speed detector and oscilloscope system (with typically 10s of GHz bandwidth) the time-lens approach has been successfully applied to measure temporal instabilities in both modulation instability \cite{Narhi-2016} and optical turbulence \cite{Suret-2016}.  The time lens idea can be combined with spectral interference with a reference field (holography) to also yield intensity and phase information \cite{Tikan-2018}.

In our recent work, we have combined the DFT and time-lens methods to simultaneously characterize the instabilities observed during start-up dynamics in a passively modelocked fiber laser, and the results in Fig.~2 illustrate typical results that have been obtained by recording real-time spectra and temporal intensity profiles over 100s of cavity roundtrip times after the laser pump diode was switched on during a phase of Q-switched mode-locked operation prior to stabilization. Full details are to found in Ref.~\cite{Ryczkowski-2018}. The results show: (a) the emergence of a 2-soliton state where the two components separate then merge before sudden decay and (b) a 3-soliton state where the components evolve independently before decay. Left and right columns show respectively the measured spectra and temporal intensity profiles respectively.  The coherence between the soliton structures seen in the time domain is manifested in the clear modulation in the measured fringes.  These results are remarkable in revealing the highly complex nature of dissipative soliton interactions that occur in modelocked lasers, and which have not been possible to observe directly before the development of real-time techniques such as DFT and the time-lens \cite{Grelu-2012}.  

\section{Classical and Quantum Imaging}
\label{sec:quantum imaging}

The previous section described how the space-time analogy first introduced in the 1970s \cite{Froehly-1973} has been applied to the field of ultrafast optics. But it has also been extremely influential in extending and developing several ideas in classical and quantum spatial imaging into new domains.   
The basis of Fourier optics consists of applying to a spatial image, concepts originally developed for temporal signals. This includes, for example, developing the formalism describing a  transfer function of spatial frequencies instead of temporal frequencies.  It means that patterns in an image can be considered as a  spatial signal, in contrast (not in contradiction) with the point of view where an image conveys a great number of (temporal) channels in parallel.  We present in the following some milestones of this approach, by strengthening, when relevant, the differences with the alternative point of view described above. 


\subsection{Parametric image amplification}
An image of a transparent object is formed on a $\chi^{(2)}$ crystal by light issued from a laser and interacts in the crystal with a collimated pump formed by the second harmonic of the same laser. Parametric down-conversion of the pump photons leads to image amplification,  provided that phase matching is fulfilled in the crystal. The amplified image is detected with a CCD camera.  We have shown \cite{devaux_transfer_1995} that a simple crystal rotation is enough to pass, for the spatial frequencies, from a low-pass amplifier to a band-pass amplifier. In this latter configuration, the edges of the object are the most amplified. For short pulses, the image, formed by the idler, is generated only during the interaction with the pump, resulting in time-gating properties: see \cite{lantz_parametric_2008} for a review.

\subsection{Spatially noiseless image amplification}

If both the signal and idler are injected in the crystal, the amplification is noiseless, i.e. preserves the signal to noise ratio because of the absence of a quantum noise channel at the crystal input. We have shown in \cite{mosset_spatially_2005} that pure spatial fluctuations recorded by a camera do obey noiseless amplification, in contrast with \cite{choi_noiseless_1999}, which showed that the signal-to-noise ratio is preserved on the image, where the considered noise was temporal, recorded by a photodiode successively at the different points of the image. This is a first clear illustration of the difference between the two points of view summarized above.

\subsection{Einstein-Podolsky-Rosen (EPR) paradox in single pairs of images}
Properties of fluctuations in quantum mechanics are described by ensemble averages, which are often estimated by time averages if the signal is stationary in time, but which can also be estimated by spatial averages if the signal is stationary in space on a sufficiently large area. Most of the experiments in quantum imaging record averages of temporal coincidences, i.e. characterize the spatial distribution of temporal averages, rather than spatial averages, in agreement with the point of view considering an image as an ensemble of information channels \cite{howell_realization_2004}. In \cite{lantz_einstein-podolsky-rosen_2015}, we have demonstrated that an Einstein-Podolsky-Rosen (EPR) paradox results from strong spatial correlations between photons of a pair in both the image and the Fourier plane by using only one pair of images in each plane. We can conclude that an image conveys quantum information by itself, without any need to employ a set of images: the ensemble average is obtained by ”repeating the experiment” over the different resolution cells, or spatial modes, of the image.
This study uses directly the concepts of Fourier optics. Indeed, the transverse momentum $p_x$ of a photon is directly recorded in the Fourier plane, since
$p_x = \hbar k_x=h \nu_x=h \, x / (\lambda f)$, where $h$ is the Planck constant, $k_x$ a transverse coordinate of the wave vector, $f$ the focal length and $x$ a coordinate in the Fourier plane, as in (\ref{fourierplane}). If $X$ is, as in this equation,  the corresponding coordinate in the object plane , the Heisenberg principle $\Delta_X \Delta p_x> \frac{\hbar}{2}$ can then be translated in Fourier optics as $2 \pi\Delta_X \Delta \nu_x > \frac{1}{2}$.  This latter relation is well known in signal processing as the minimum product of the extensions of a signal in the direct and the Fourier space \cite{Cohen_1989}. Hence, the Heisenberg principle can be viewed in optics as the conjunction of diffraction laws of wave optics, i.e. the duality between the resolution in the  direct and Fourier space,  with the corpuscular nature of light. The EPR paradox means that correlations due to entanglement allow  an apparent violation of this principle, when considering the conditional variances.

\subsection{Temporal ghost imaging by image cross-correlations}
Ghost imaging (GI), whether quantum or classical, is an archetypal example of the second point of view developed above. In the reference arm, a channel of the image (a pixel) is recorded with temporal resolution by a  photodiode, without interaction with the object.  Crucially, the temporal signal is random, either because of its quantum nature or because of classical fluctuations, for example of a (pseudo) thermal signal. In the test arm, the different channels corresponding to the different pixels are transmitted by the object and summed together on a bucket detector, without spatial resolution but with temporal resolution. The value of the object for a pixel is retrieved by cross-correlation of the reference and test signals, thanks to the identical temporal fluctuations of the reference channel and of the part of the test channel that is associated with the considered pixel. The only difference between quantum GI \cite{pittman_optical_1995} and thermal GI \cite{ferri_high-resolution_2005} is that in the first case the identity of fluctuations is due to twin photons while, in the second case, the fluctuating beam is simply divided in two by a beam-splitter, or computed in the case of computational GI \cite{shapiro_computational_2008}.

Considering an image as a fluctuating signal leads naturally to temporal GI. The “object” to retrieve is now a temporal transparency, with different levels succeeding in time. On the reference arm, successive random images are recorded, or computed, with temporal and spatial resolution. On the test arm, each image is multiplied by the corresponding temporal level of the object and summed with the other images on a camera, with spatial resolution but no temporal resolution. The temporal object is retrieved by cross-correlation of the successive images with the summed image. Note the exact space-time transposition between GI and temporal GI. We have demonstrated computational temporal GI \cite{devaux_computational_2016}, temporal GI with  speckle images \cite{devaux_temporal_2016} and quantum temporal GI \cite{denis_temporal_2017}. 

Temporal GI refers also to GI entirely in the temporal domain \cite{ryczkowski_ghost_2016}. In the test arm, the light is transmitted through the “time object” and detected with a slow single pixel detector (SPD) that cannot properly resolve the time object. In the reference arm, the light that does not interact with the temporal object is detected with a fast SPD. This scheme and the scheme developed in \cite{devaux_computational_2016,devaux_temporal_2016,denis_temporal_2017} have each their own advantages and drawbacks. The main advantage of the latter consists of the replacement of thousands of synchronized replicas of the temporal signal required in \cite{ryczkowski_ghost_2016} by the use of a detector array with thousands of pixels (the camera). Its drawback is the slowness (at most 27 Khz in \cite{devaux_temporal_2016}) that prevents its use to cancel dispersion effects as developed in \cite{ryczkowski_ghost_2016}.
\subsection{Conclusion of this section}
The works reported in this section result from the crossing of the local tradition of Fourier optics with recent developments in nonlinear and quantum imaging (note nevertheless that classical GI is pure linear imaging).  In particular, the space-time transpositions have been systematically developed, resulting for example in the notion of spatial coincidences detected with single photon sensitive cameras, as an alternative to temporal coincidences detected with time-gated fast detectors. Since the image is considered as  a spatial signal, the  bi-dimensional Fourier transform is present in all stages of the studies. Some examples are the following.  A split-step  algorithm is used to simulate the effect of diffraction and of the nonlinear interaction in the crystal. At each step, the non linear interaction is taken into account in the direct space, while a Fourier transform in the transverse plane allows  the  propagation of the plane wave spectra. Results extend from the calculation of  the transfer function due to phase matching to the computation of quantum correlations that build in the crystal. On the experimental side, all cross-correlations of twin images are calculated by multiplications in the Fourier domain.

\section{Optical computing and photonic neural networks}
\label{sec:PhotNN}
\begin{figure}[t]
	\includegraphics[width=0.5\textwidth]{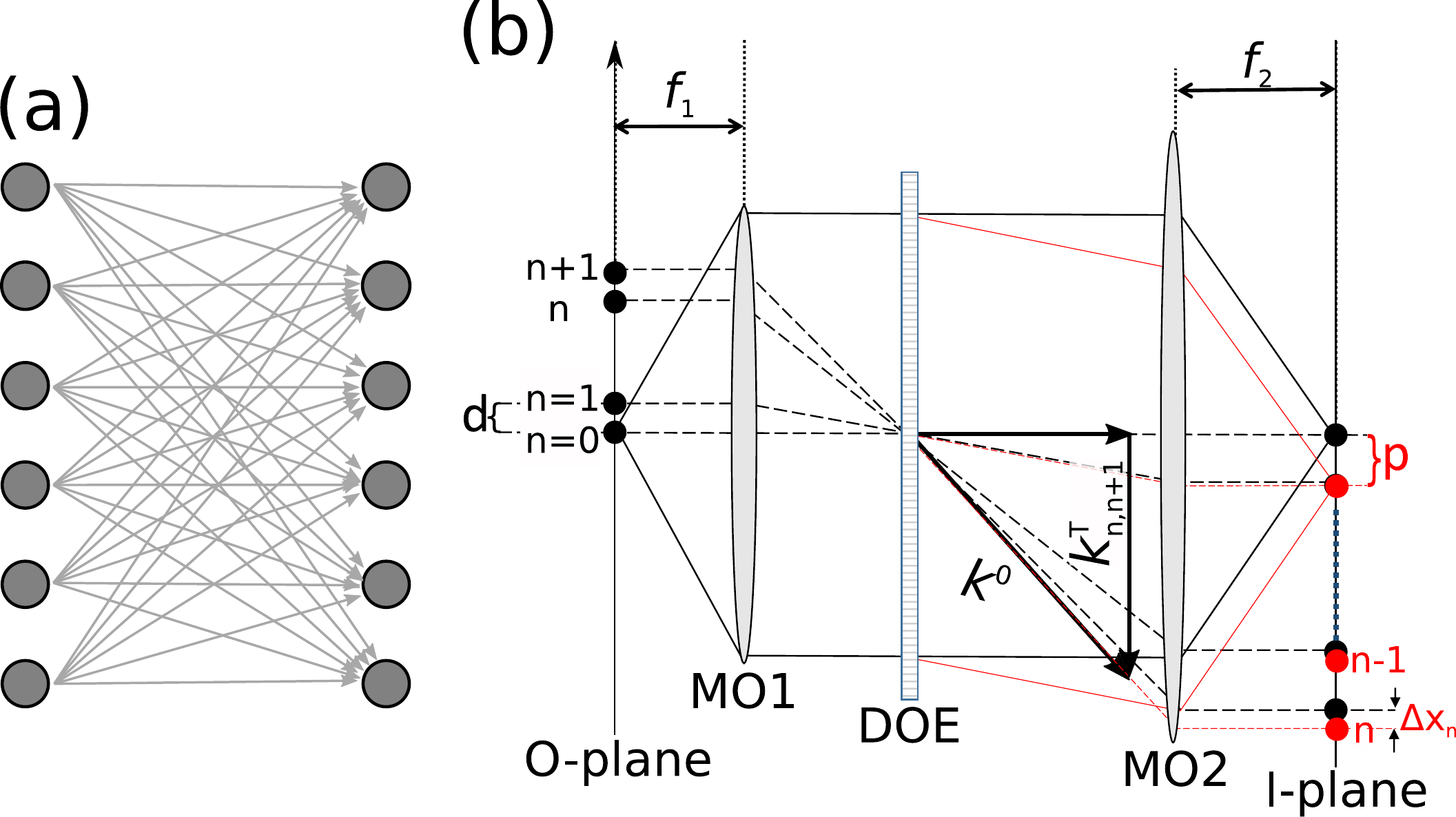}
	\caption{\label{fig:NNdiff}
		(a) Two layers of neurons (circles) connected into a neural network.
		The state of neurons in the left layer serves as input for the layer on the right.
		(b) Similar coupling can be implemented in a telescope of two infinity corrected microscope objectives (MO1\&2), where a diffractive optical element (DOE) located in the beam-path creates the connections in parallel.
		This system was used to demonstrate learning in the context of a recurrent neural network \cite{Bueno-2018}.}
	\vspace{-0.5cm}
\end{figure} 

Photonic implementations of neural networks have experienced a revival in recent years \cite{VanderSande-2017}.
 While the exploding interest in neural networks was stimulated by novel applications and concepts \cite{LeCun-2015}, the motivation for optical neural networks remains the one of the original efforts in the 1980s \cite{Farhat-1985}: an advantage of photonics in terms of parallelism and signal transmission when compared to electronics \cite{Lohmann-1990}.
 
 Photonic systems have recently revolutionized the hardware implementation of Recurrent Neural Networks and Reservoir Computing \cite{Jaeger-2004}, in particular. The fundamental principles of Reservoir Computing strongly facilitate a realization in such complex analog systems. Especially delay systems \cite{duvernoy1987bistabilite, larger2010}, which potentially provide numerous degrees of freedom, can efficiently be exploited for information processing applied to photonic neural networks \cite{martinenghi2012, larger2017}. 
 
 We also demonstrated other architectures that consider a neural network's fundamental features, see Fig. \ref{fig:NNdiff}(a), taking into account the potential of photonics and the relevance of Fourier optics.
 A neural network's purpose is to compute, which corresponds to nonlinearly mapping input information onto the desired result.
 For most tasks this mapping is non-trivial, yet it can be decomposed into a series of simple nonlinearities weighted by coefficients - much like in a polynomial series approximation.
 Neurons provide the simple nonlinearities, while the scaling-coefficients correspond to weighed connections which define the network's topology.
 In principle, nonlinearities can be scaled and summed in parallel, and accordingly neural networks are parallel computing concepts.

Many nonlinear photonic components can serve as a neuron in principal.
 The challenge is realizing parallel network connections, which is where the advantage of the non-interacting photons comes into play.
 A simple telescope of two microscope objectives (MO1, MO2), schematically illustrated in Fig. \ref{fig:NNdiff}(b), images neurons located in the object plane ($O$) onto the image plane ($I$). 
 Within the collimated space between MO1 and MO2, the $n$th neuron's planar wave front is identified by transversal wavevector $k^{T}_{n}=k^{0} x^{O}_{n} / f_{1}$, with $x^{O}_{n} = n d$.
 Focusing by MO2 creates the photonic neuron's image at $x^{I}_{n} = f_{2} k^{T}_{n} / k^{0}$.
 Here, $f_{1}=20~$mm and $f_{2}=18~$mm are the focal distance of MO1 and MO2, respectively.
 Coupling neuron $n$ to neuron $m$ requires their fields to spatially overlap in plane $I$, hence a coupling mechanism incorporated between MO1 and MO2 needs to contribute $\Delta k^{T}_{n,m} = k^{T}_{m} - k^{T}_{n}$.
 Since emission of photonic neurons traverses the optical system simultaneously without interaction, coupling is established fully in parallel.

Early experiments with photonic Hopfield networks created the $\Delta k^{T}$s optimized for a specific task using volume holograms \cite{Farhat-1985}.
 The holographic coupling concept can in principle be extended to deep neural network trained by gradient back-propagation \cite{Wagner-1987}.
 For that, the phase-conjugated version of the desired output is propagated through the optical system in a backward direction, writing the hologram of the coupling matrix required in the forward direction.
 However, it was realized that these powerful approaches were not scalable in terms of energy-cost and system size.
 
\begin{figure}[t]
	\includegraphics[width=0.4\textwidth]{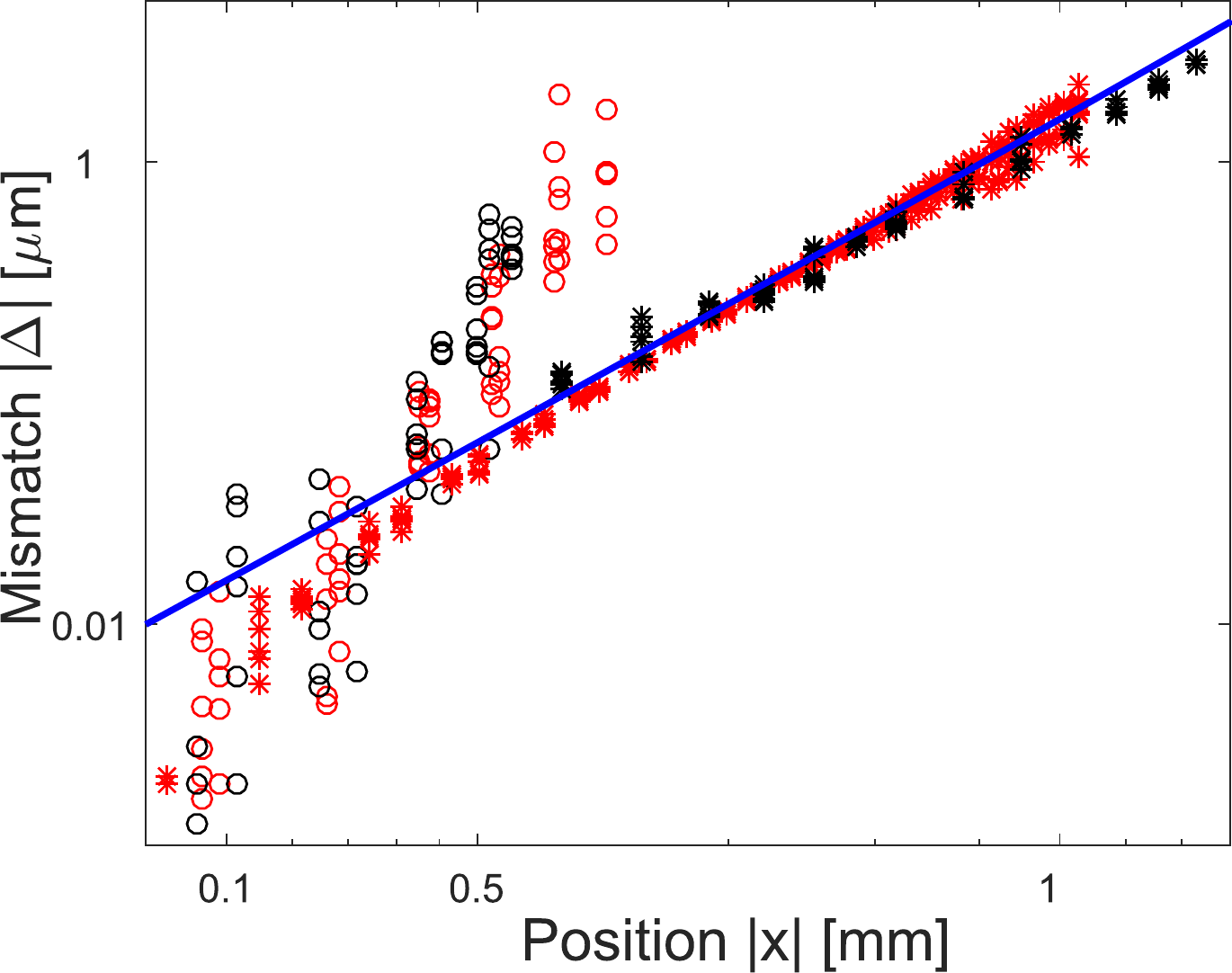}
	\caption{\label{fig:CiffCplSimExp}
		Coupling mismatch $\Delta x_n$ in diffractively coupled photonic networks.
		Analytical (blue line), experimental (black symbols) and numerical simulations (red symbols) excellently agree.
		Red circles are based on numerical simulations with identical microscope objectives as in the experiment (black circles).
		The red stars are based on numerical simulations replacing MO1 with a higher numerical aperture model with identical magnification \cite{Maktoobi2020}.
		Black stars are experimental results based on rotating the DOE, hence avoiding beam vignetting.}
	\vspace{-0.5cm}
\end{figure}

Recently, the reservoir computing concept enables using a complex, yet not task-specifically optimized nonlinear network.
 The optimization during learning is restricted to the final connection matrix, which strongly reduces the complexity for hardware implementations \cite{Jaeger-2004}.
 We demonstrated that such coupling can be created by a single, off-the-shelf diffractive optical element (DOE) \cite{Bueno-2018}.
 In its action, the DOE is a diffraction grating with a particular intensity distribution across diffractive orders.
 It is therefore a periodic phase modulation which creates the $\Delta k^{T}$s. However, this diffraction based technique results in a continuously increasing mismatch $\Delta x_{n}$ between $x^{I}_{n}$ and the ideal position $n d f_{2} / f_{1}$, see Fig. \ref{fig:NNdiff}(b).
 Using $B = \frac{n d}{\sqrt{{f_1}^2+n^2  d^2}} + \frac{\lambda}{p^{DOE}}$, we find
\begin{equation} \label{eq:Delta}
\Delta x_{n}= \frac{f_{2}}{f_{1}}(n+1)d - \frac{ B f_2}{\sqrt{1- B^2}},
\end{equation}\\
\noindent where in our experiment $p^{DOE}=0.9~$mm and $\lambda = 662.1~$nm.
 The blue line in Fig. \ref{fig:CiffCplSimExp} shows that mismatch $\Delta x_{n}$ for an emitter position $x^{O}_{n} < 5~$mm remains below 1$~\mu$m.
 The mismatch is therefore of the order of a single mode emitter's radius, corresponding to the limit beyond which coupling is lost.
\vspace{+0.5cm}

A more complete evaluation of networks formed by diffraction requires a model for the propagating optical fields with a sensitivity and accuracy of $\Delta x_{n}\leq 1~\mu$m for $x^{O}_{n}>5~$mm.
 Considering the typical focal distance of microscope objectives this rules out paraxial approximations, and propagation between MO1 and MO2 requires the non-paraxial propagation of the planar wave spectrum $A(\nu_{x},\nu_{y})$ \cite{Goodman-2017}.
 Here, $\nu_{x}$ and $\nu_{y}$ are the spatial frequencies of the planar wave spectrum.
 Propagating from one plane to the next is obtained by Fourier transformations ($\mathcal{F}$) of the field, with a phase shift according to $\exp\left(i2\pi z \sqrt{\frac{1}{\lambda^2} - v_x^2-v_y^2}\right)$.
 Beam propagation between MO1 and MO2 along the z-direction is computed according to:
 
\begin{equation} \label{eu_eqn}
A(x,y,z)=\mathcal{F}\left\lbrace A(\nu_{x},\nu_{y})\exp\left(i2\pi z \sqrt{\frac{1}{\lambda^2} - v_x^2-v_y^2}\right)   \right\rbrace 
\end{equation}
 The DOE's effect is included as an additional phase modulation, while propagation through the high numerical aperture microscope objectives is efficiently realized with the Debye integral method \cite{Born-2013,Wolf-1981,Sheppard-2000,Leutenegger-2006}.
 Crucially, the Debye integral relates a microscope objective's image plane coordinates to spatial frequencies according to $x^{I}|y^{I}=\lambda f_{2}\nu_{x|y}$. This corresponds to the paraxial approximation, and a correction according to $x^{I}|y^{I} = (\lambda f_{2}  \nu_{x|y}) / {(\sqrt{1-\lambda^{2} (\nu_{x|y})^{2} }}$ is required.
 
Black data in Fig. \ref{fig:CiffCplSimExp} show experimental results of the diffractive coupling scheme, which agree exceptionally well with numerical simulations (red data) based on identical parameters.
 Two different systems were compared.
 Using a low NA MO1 (NA<0.2, black circles in Fig. \ref{fig:CiffCplSimExp}) causes beam vignetting during collimation, and the resulting diffraction at the clear aperture's edge causes strong deviation from the analytical expression of $| \Delta x_{n} |$, blue line in Fig. \ref{fig:CiffCplSimExp}.
 This effect is correctly described by the numerical algorithm, see red circles.
By tilting the DOE we emulated collimated beam angles of photonic neurons at different positions, crucially avoiding beam vignetting as the collimated beam remains centered, and experimental results are shown as black stars.
 Replacing MO1 with a high NA, low magnification microscope objective removes the vignetting effect for photonic neurons inside a much larger area, as confirmed by numerical simulations (red stars).
 Numerically and experimentally obtained data for systems avoiding vignetting perfectly follows the analytical description.
 Non-paraxial Fourier techniques therefore very accurately describe diffractive photonic networks spanning areas of $\sim 25~$mm$^2$, providing a resolution and sensitivity significantly better than $1~\mu$m.
 Considering a realistic spacing of $10~\mu$m between photonic emitters, this confirms diffractive coupling concepts for networks hosting approximately half a million optical nodes.

We have demonstrated diffractive networks consisting of semiconductor lasers \cite{Brunner-2015} and the pixels of a SLM \cite{Bueno-2018}.
 The latter system was amended with a digital micro-mirror device which implemented adjustable readout weights.
 Combined with a evolutionary learning algorithm, we trained a network of 900 photonic neurons to perform the prediction of a chaotic sequence.
 The possibly large network size and low complexity makes diffractive coupling an excellent technique for a wide range of systems and applications.

\section{Conclusions}

The widespread application of Fourier concepts in optics is not surprising from a modern perspective, but it is difficult to imagine a pioneer such as Duffieux envisaging their use in fields so removed from traditional image formation.  Many of the areas discussed above are finding significant applications in a remarkable range of different fields.  For example, the ability to shape and control the spatial structure of high-power femtosecond Bessel beams is being actively applied to important uses in material processing, and spatial shaping concepts with purely linear optics are opening up new possibilities in the design of new paradigms in optical artificial intelligence. The real-time measurement technique of the DFT and the time lens are being used to gain new insights into several different classes of optical instability, including the emergence of extreme events in systems other than optics - such as the celebrated rogue waves on the ocean and in hydrodynamics \cite{Dudley-2014}.We hope that the results described above have clearly shown how a clear physical understanding of Fourier optics can lead to significant results that touch on areas of photonics and optics that are both relatively well established (such as Bessel beam shaping) to those that have appeared only recently (such as in photonic artificial intelligence). Of course the underlying principles of our experiments have been understood for decades, and it has rather been the recent availability of new technologies in areas such as spatial light modulators that have allowed their practical implementation.  Indeed, as experimental developments continue, there is every reason to expect that even more unexpected applications will emerge.  But whilst research must always be forward-looking and seek new applications for the future, it is our hope that the work of those that paved the way will not be forgotten. 

\section{Acknowledgments}
We would like to acknowledge discussions and valuable contributions from a number of co-workers, especially: G. Genty, P. Ryczkowski, C. Billet, and L. Furfaro. 

\section{Funding Information}
We acknowledge support from: the EIPHI Graduate School (ANR-17-EURE-0002);R\'egion Bourgogne  Franche-Comt\'e; European Research Council (ERC) (682032-PULSAR); Volkswagen Foundation (NeuroQNet); French Investissements d'Avenir programme, project ISITE-BFC (contract ANR-15-IDEX-0003).  

\bibliography{Biblio_Duffieux}


\end{document}